\documentclass[amsmath,amsfonts,amssymb,aps,prc,superscriptaddress,preprint,preprintnumbers]{revtex4-1}
\usepackage{natbib}

\newcommand{\bra}[1]{\left\langle #1 \right|}
\newcommand{\ket}[1]{\left|#1\right\rangle}
\newcommand{\braket}[2]{\left\langle#1 |  #2\right\rangle}
\newcommand{\brakets}[3]{\left\langle#1 |\, #2\, | #3\right\rangle}

\newcommand{\tee}{(t)}
\newcommand{\AdS}{\mathrm{AdS}}

\numberwithin{equation}{section}

\begin{document}
\title{Conformal Blocks for the 4-Point Function in Conformal Quantum Mechanics}
\author{R. Jackiw}
\affiliation{Department of Physics,
MIT, 
Cambridge, MA 02139}
\author{S.-Y. Pi}
\affiliation{
Department of Physics,
Boston University,
Boston, MA 02215}

\begin{abstract}
Extending  previous work on 2 -- and 3 -- point functions, we study the 4 -- point function and its conformal block structure in conformal quantum mechanics CFT$_1$, which realizes the $SO(2,1)$ symmetry group. Conformal covariance is preserved even though the operators with which we work need not be primary and the states are not conformally invariant.  We find that only one conformal block contributes to the four-point function. We describe some further properties of the states that we use and  we construct dynamical evolution generated by the compact generator of $SO(2.1)$.
\end{abstract}

\pacs{}
\preprint{MIT-CTP/4365}
\maketitle

\section{Introduction and Review}
A recent Letter [1] initiated research on the $\AdS_{d + 1}/\mathrm{CFT}_d$ correspondence for the special case $d = 1$. This dimension corresponds to the lowest ``rung"on the dimensional ``ladder" of $SO(d + 1, 1)$ conformally invariant scalar field theories in $d$ dimensions.
\begin{equation}
\mathcal{L}_d = \frac{1}{2}\ \partial_\mu \, \Phi \, \partial^\mu\, \Phi - g \, \Phi^{\frac{2d}{d-2}}
\label{rjAdS1}
\end{equation}
At $d=1\ [\Phi \left(t, {\bf r}) \to q (t)\right)]\ \mathcal{L}_1$ governs conformal quantum mechanics with a $g/q^2$ potential \cite{Jackiw:1972ph}, and supports an $SO (2,1)$ symmetry, with generators $H, D$ and $K$.

\noindent
Their algebra
\begin{subequations}\label{rjAdS2}
\begin{eqnarray}
i [D, H] &=& H \label{rjAdS2-a},\\
i [D, K] &=& - K \label{rjAdS2-b},\\
i [K, H] &=& 2D \label{rjAdS2-c},
\end{eqnarray}
\end{subequations}
when presented in Cartan basis, 
\begin{subequations}\label{rjAdS3}
\begin{eqnarray}
R \equiv \frac{1}{2} \left(\frac{K}{a} + a\, H \right) \label{rjAdS3-a},\\
L_\pm \equiv \frac{1}{2}\, \left(\frac{K}{a} - a\, H \right)\, \pm i\, D,  \label{rjAdS3-b}
\end{eqnarray}
\end{subequations}
reads
\begin{subequations}\label{rjAdS4}
\begin{eqnarray}
[R, L_\pm] &=& \pm\, L_\pm , \label{rjAdS4-a}\\
\left[L_{-} , L_{+}\right] &=& 2\, R\, . \label{rjAdS4-b}
\end{eqnarray}
\end{subequations}
($a$ is a scaling parameter with dimension of time; frequently we set it to 1.)

In spite of the natural position that $d=1$ enjoys, various questions arise about the correspondence. $\AdS_2$ calculations allegedly produce boundary $N$-point correlation functions in CFT$_1$.
\begin{equation}
G_N \, (t_1, \ldots , t_N) \, \sim \, \langle \varphi_1 (t_1) \ldots \varphi_N (t_N)\rangle 
\label{rjAdS5}
\end{equation}
where $\varphi (t)$ are primary operators in the boundary conformal theory, and the averaging state $\langle \ldots \rangle $ is conformally invariant, {\it i.e.} it is annihilated by the conformal generators. However, in $\mathrm{CFT}_1$ normalized states are not invariant and invariant states are not normalizable, rendering problematic calculation of expectation values. Furthermore, one wonders which operators in conformal quantum mechanics realize the primary operators $\varphi (t)$, whose correlation functions arise from the $\AdS_2$ calculation.

These puzzles are resolved in the Letter [1]. We focus on the $R$ operator, taken to be positive $(g \geqslant 0)$ and defined on the half-line $(q \geqslant 0)$, with integer-spaced eigenvalues $r_n$ and orthonomal eigenstates $| n \rangle$.
\begin{subequations}\label{rjAdS6}
\begin{gather}
R \ket{n} = r_n\, \ket{n}\label{rjAds6-a}\\
r_n = r_0 + n,\ \  r_0 > 0, \ \ n = 0,1 \ldots \nonumber\\
\braket{n}{n^\prime} = \delta_{n\, n^\prime} \nonumber \\
L_\pm \ket{n} = \sqrt{r_n \, (r_n \pm 1) - r_0 \, (r_0 -1)} \, \ket{n \pm 1} \label{rjAds6-b}
\end{gather}
\end{subequations}

We need states that carry a representation of the $SO(2,1)$ action. To this end we constructed the operator $\mathcal{O}\tee$,
\begin{eqnarray}
\mathcal{O}\tee &=& N\tee\, \mathrm{exp} - \left(\omega\tee\, L_+\right),\nonumber\\[1.5ex]
N\tee &=& \bigg[\Gamma (2r_0)\bigg]^{\frac{1}{2}}\ \left[\frac{\omega\tee + 1}{2}\right]^{2 r_0}, \nonumber \\[.75ex]
\omega\tee &=& \frac{a + i\, t}{a - i\, t} = e^{i \theta}\ \ \mbox{where}\ t = a \tan \theta/2 ,
\label{rjAdS7}
\end{eqnarray}
and defined ``$t$ states" $\ket{t}$ by the action of $\mathcal{O}\tee$ on the $R$-vacuum.
\begin{eqnarray}
\ket{t} = \mathcal{O}\tee\, \ket{n=0}  \label{rjAdS8}\\
R\, \ket{n=0} = r_0\, \ket{n=0}\label{rjAdS9}
\end{eqnarray}
From their definition \eqref{rjAdS8} it follows that the $\ket{t}$ states satisfy \cite{rjbibitem3}

\begin{subequations}\label{rjAdS10}
\begin{eqnarray}
H\ket{t} &=& - i\, \frac{d}{dt} \, \ket{t}, \label{rjAdS10-a}\\[.5ex]
D \ket{t} &=& - i\, \left(t\, \frac{d}{d t} + r_0 \right)  \ket{t}, \label{rjAdS10-b}\\[1ex]
K \ket{t} &=& -i \left(t^2\,  \frac{d}{d t} + 2\, r_0\, t \right) \ket{t}. \label{rjAdS10-c}
\end{eqnarray}
\end{subequations}

 $N$-point functions are constructed from the $\ket{t}$ states. For $G_N (t_1, \ldots , t_N)$,  the averaging state $\langle \ldots \rangle$ is the $R$-vacuum $\ket{n = 0}$.
The first and last operators are taken to be $\mathcal{O}^\dagger (t_1)$ and $\mathcal{O} (t_N)$, while the remaining $N-2$ operators are conventional but unspecified primary operators $\varphi$, with scale dimension $\delta$. 
\begin{subequations}\label{rjAdS11}
\begin{eqnarray}
i [H, \varphi\tee] &=& \frac{d}{d t}\ \varphi \tee  \label{rjAdS11-a}\\[1ex]
i [D, \varphi\tee] &=& \left(t\, \frac{d}{dt}\ + \delta\right)\ \varphi\tee  \label{rjAdS11-b}\\[1ex]
i [K, \varphi\tee] &=& \left(t^2\,  \frac{d}{dt}\ + 2\, \delta t\right)\ \varphi \tee  \label{rjAdS11-c}
\end{eqnarray}
\end{subequations}
Thus an $N$-point function involves the $\ket{t}$ states.
\begin{align}
&G_N\, (t_1, t_2, \ldots , \ t_{N-1}, t_N) =\nonumber \\
&\bra{n=0} \, \mathcal{O}^\dagger (t_1)\, \varphi_2\, (t_2) \dots \varphi_{N-1}\, (t_{N-1})\ \mathcal{O} \, (t_N) \ket{n=0}\label{rjAdS12}\\
&= \brakets{t_1}{\varphi_2 (t_2)\ldots \varphi_{N-1}\, (t_{N-1})}{t_N} \nonumber
\end{align}

In spite of the fact that the $\mathcal{O}{\tee}$ operators are not primary, and the averaging state $\ket{n=0}$ is not conformally invariant, the two ``defects" cancel and the resultant $N$-point functions satisfy conformal covariance conditions. Consequently, in an operator-state correspondence we may consider the operators $\mathcal{O} {\tee}$, when acting on the states $\ket{n=0}$, as primary with dimension $r_0$.


In this way one establishes that \cite{rjbibitem3-a}, \cite{rjbibitem3a}
\begin{eqnarray}
G_2 (t_1, t_2) &=& \braket{t_1}{t_2} = \brakets{n=0}{\mathcal{O}^\dagger (t_1)\, \mathcal{O} (t_2)}{n=0} \nonumber\\[1ex]
&=& \frac{\Gamma\, (2 r_0)\, a^{2 r_0}}{[2 i\, (t_1 - t_2)]^{2 r_0}} ,
\label{rjAdS13}
\end{eqnarray}
\begin{gather}
G_3 \, (t_1, t, t_2) = \brakets{t_1}{\varphi \tee}{t_2} = \brakets{n=0}{\mathcal{O}^\dagger\, (t_1)\, \varphi (t)\, \mathcal{O} (t_2)}{n= 0} \nonumber \\[1ex]
= \brakets{n=0}{\varphi (0)}{n=0} \, \left(\frac{i}{2}\right)^{2 r_0 + \delta}  \frac{\Gamma\, (2 r_0)\, a^{2 r_0}}{(t_1 - t)^\delta (t-t_2)^\delta (t_2 - t_1)^{2 r_0 -\delta}}\  .
\label{rjAdS14}
\end{gather}
 The expressions \eqref{rjAdS13}, \eqref{rjAdS14} also arise from calculations based on a scalar field in $\AdS_2$, at the boundary of the $\AdS_2$ bulk.

In Section \ref{rjSec2}, we extend the investigation to the quantum mechanical 4-point function.
\begin{gather}
G_4 \, (t_1, t_2, t_3, t_4) = \brakets{t_1}{\varphi (t_2)\, \varphi (t_3)}{t_4}\nonumber\\ 
= \brakets{n=0}{\mathcal{O}^\dagger\, (t_1)\, \varphi (t_2)\, \varphi (t_3)\, \mathcal{O} (t_4)}{n= 0}
\label{rjAdS15}
\end{gather}
The two $\varphi$ fields are taken to be identical, with scale dimension $\delta$. We demonstrate that conformal covariance and block structure are maintained by our unconventional realization of the conformal symmetry: once again ``defects" cancel.

In Section \ref{rjSec3}, we study some further properties of the $\ket{t}$ states and of related energy eigenstates $\ket{E}$ of the Hamiltonian $H$. Also we show how the $R$ operator can replace $H$ as the evolution generator.

\section{Correlation Function and Conformal Block}\label{rjSec2}
\subsection{4-point Function in CFT$_1$}\label{rjSec2-A}
To calculate $G_4$ in \eqref{rjAdS15}, insert complete sets of $\ket{n}$ states between the operators. Also without loss of generality evaluate the sums at special values: $t_1 = -i a, t_4 = i a$. [This may always be achieved by a complex $SO(2,1)$ transformation.] One is left with a single sum. It remains to reduce matrix elements $\brakets{n}{\varphi\tee}{n^\prime}$ to $\brakets{n=0}{\varphi (0)}{n^\prime = 0}$. This was accomplished by dAFF \cite{Jackiw:1972ph} with the $SO(2,1)$ Wigner-Eckart theorem. This procedure leads to\cite{rjbibitem4}
\begin{eqnarray}
G_4 \, (t_1, t_2, t_3, t_4) &=& \mid \brakets{n=0}{\varphi (0)}{n=0} \mid^2\, \frac{\Gamma^2 (1-\delta)}{2^{2 \delta + 2 \, r_0} }\nonumber\\[1ex]
\times \  \frac{\Gamma^2 (2 r_0)}{(t_{1 3}\, t_{2 4})^{2\delta}\, (t_{1 4})^{2 r_0 - 2\delta}} &\ \sum\limits^\infty_{n=0}\ & \frac{1}{\Gamma(2 r_0 + n)\, \Gamma^2 \, (1-\delta - n)}\ \frac{x^{n-\delta}}{n !} , \nonumber\\[1ex]
t_{ij} \equiv t_i - t_j , \ \ x &\equiv& \frac{t_{12}\, t_{34}}{t_{13} \, t_{24}}\,  .
\label{rjAdS2-1}
\end{eqnarray}
(The scaling parameter $a$ is set to unity.)

Remarkably, the sum may be evaluated in terms of the hypergeometric function $_2 F_1$. The final expression for $G_4$ is 
\begin{gather}
G_4 (t_1, t_2, t_3, t_4) = |\brakets{n=0}{\varphi (0)}{n=0}|^2\, \frac{1}{2^{2\delta + 2 r_0}}\nonumber\\[1ex]
\times \quad \frac{\Gamma (2 r_0) \ x^{r_0} \,  _2F_1 \ (\delta, \delta; 2 r_{0}; x )}{(t_{1 3}\, t_{2 4})^{\delta - r_0} \ (t_{12}\, t_{34})^{\delta + r_0}\ (t_{14})^{2 r_0 - 2 \delta}} \ .
\label{rjAdS2-2}
\end{gather}
The polynomial in $t_{ij}$ provides conformal covariance, while the $x$-dependence is conformally invariant. (In one dimension four points lead to a single invariant, as opposed to two invariants in higher dimensions.)

The 4-point function may be presented by a Mellin transform since $_2 F_1$ possesses a Mellin-Barnes representation.
\begin{equation}
_2F_1 \ (\delta, \delta; 2 r_{0}; x ) = \frac{\Gamma(2 r_0)}{\Gamma^2 (\delta)}\ \int\limits^{i \infty}_{-i \infty} \, d s \ \frac{\Gamma^2 (\delta + s) \, \Gamma (-s)}{\Gamma\, (2 r_0 + s)}\ (-x)^s
\label{rjAdS2-3}
\end{equation}
The sum in \eqref{rjAdS2-1} arises from the poles of $\Gamma (- s)$ in \eqref{rjAdS2-3}. A single Mellin integral suffices at $d=1$ because there is only a single invariant.

\subsection{Conformal Block in CFT$_{\mathbf{1}}$}\label{rjSec2b}
\noindent In general one expects that the 4-point function $G_4$ may be presented as a superposition of ``conformal blocks." These quantities are kinematically determined by the eigenfunctions of the $SO(2,1)$ Casimir. This is like a partial wave expansion of a scattering amplitude --- indeed ``conformal partial waves" is an alternative nomenclature.

Conformal blocks at arbitrary $d$ for $SO(d+1, 1)$ have been extensively studied by Dolan and Osborn. Recently they have constructed the $d=1, SO(2,1)$ quantities by passing to the (somewhat singular) limit $ d \to 1$ for a block coming from a single operator and its descendants \cite{Dolan:2011dv}. In contrast, from the start we work directly with the $SO (2,1)$ symmetry at  $d = 1$. 

We present the general 4-point function.
\begin{gather}
G_4 \, (t_1, t_2, t_3, t_4) =\langle \varphi_1 \, (t_1)\, \varphi_2 (t_2)\, \varphi_3 (t_3)\, \varphi_4 (t_4)\rangle  \nonumber \\[1ex]
= \frac{1}{(t_{12})^{\Delta_1 +\Delta_2}\, 
(t_{34})^{\Delta_3 + \Delta_4}\, 
(t_{13})^{\Delta_{3 4}}\, 
(t_{14})^{\Delta_{12} - \Delta_{34}}\,
(t_{24})^{-\Delta_{12}}   }  \ \ F (x) \nonumber \\[1ex]
= p (t_1, t_2, t_3, t_4) \ F (x)
\label{rjAdS2-4}
\end{gather}
The $t$-polynomial $p$ carries the conformal transformation property of $G_4$, while $F$ is invariant. $\Delta_i$ is the dimension of $\varphi_i$ and $\Delta_{ij} \equiv \Delta_i - \Delta_j$. (This expression is more general than the one we used in our previous discussion, which is specialized to $\Delta_1 = \Delta_4 = r_0, \ \Delta_2 = \Delta_3 = \delta, \ \varphi_1 = \mathcal{O}^\dagger, \varphi_4 = \mathcal{O}, \, \varphi_{2, 3} = \varphi .)$

The block decomposition states
\begin{equation}
F (x) = \sum\limits_i \, b_i\, B_i (x),
\label{rjAdS2-5}
\end{equation}
where $i$ labels the kinematical variety of blocks $B_i$. Each $B_i$ is constructed from a specific primary operator and its descendants. The $b_i$'s contain dynamical data. The blocks are eigenfunctions of the Casimir.   
\begin{eqnarray}
C &=& \frac{1}{2} \ (H K + K H) - D^2 \label{rjAdS2-6}\\[1ex]
C \, (p B) &=& c\, (p B)
\label{rjAdS2-7}
\end{eqnarray}
In \eqref{rjAdS2-6}, \eqref{rjAdS2-7}, the individual generators are sums of the corresponding derivative operators
\begin{eqnarray}
H &=& H_1 + H_2, \, K = K_1 +K_2, \, D = D_1 +D_2 \nonumber\\[1ex]
H_i &=& i\, \frac{\partial}{\partial t_i}, \, D_i = i \, \left(t_i\, \frac{\partial}{\partial t_i} + \Delta_i \right), \, K_i = i \, \left(t^2_i\,  \frac{\partial}{\partial t_i} + 2 \Delta_i t_i\right).
\label{rjAdS2-8}
\end{eqnarray}
$c$ is the eigenvalue. Thus the derivative operator $\mathcal{D}$ corresponding to $C$
\begin{equation}
\mathcal{D} \equiv - t^2_{12}\ \frac{\partial^2}{\partial t_1 \, \partial t_2} + 2\, t_{12}\, \left(\Delta_2\,  \frac{\partial}{\partial t_1} - \Delta_1 \,  \frac{\partial}{\partial t_2}\right) + (\Delta_1 + \Delta_2)^2 - (\Delta_1 + \Delta_2) ,
\label{rjAdS2-9}
\end{equation}
acts on $p B$ as
\begin{equation}
\mathcal{D} \, (p B) = p \, \bigg(x^2 \, (1-x)\, B^{\prime\prime} + (-1 + \Delta_{12} - \Delta_{34})\, x^2 B^\prime + \Delta_{12} \Delta_{34}\, x\, B \bigg)
\label{rjAdS2-10}
\end{equation}
(dash signifies $\frac{d}{d x}$). The eigenvalue equation reads
\begin{equation}
x^2 \, (1-x)\, B^{\prime\prime} + (-1 + \Delta_{12} - \Delta_{34})\, x^2 \, B^\prime + \Delta_{12}\, \Delta_{34}\, x\, B= c \, B ,
\label{rjAdS2-11}
\end{equation}
and is solved by
\begin{subequations}\label{rjAdS2-12}
\begin{eqnarray}
B &=& x^\Delta\,  _2F_1\, (\Delta - \Delta_{12}, \Delta + \Delta_{34}; 2\, \Delta;  x).\label{rjAdS2-12a}\\
c &=& \Delta (\Delta - 1)
\label{rjAdS2-12b}
\end{eqnarray}
\end{subequations}
In order to match this block to the 4-point function \eqref{rjAdS2-2} where $\Delta_1 = \Delta_4 = r_0, \Delta_2 = \Delta_3 = \delta$ we must set $\Delta = r_0$, so that
\begin{equation}
B = x^{r_0} \, _2F_1 \, (\delta, \delta_; \, 2\,  r_{0}; x).
\label{rjAdS2-13}
\end{equation}
Evidently the single block \eqref{rjAdS2-13} reproduces the 4-point function. It is a surprise that one block suffices.

The usual route to conformal blocks is through the short-distance expansion for $\varphi_1\, (t_1)\, \varphi_2\, (t_2)$. In our construction $\varphi_1\, (t_1)$  is replaced by $\mathcal{O}^\dagger (t_1)$, which does not have an evident short distance expansion with $\varphi_2 (t_2)$. Nevertheless, within our approach we are able to derive a block representation for the 4-point function. This puts into evidence once again that our method, with its cancellation of ``defects," preserves conformal covariance.

\subsection{Generalization}\label{rjSec2c}
We have evaluated $\brakets{t_1}{\varphi (t_2)\, \varphi (t_3)}{t_4} $
where the two $\varphi$ fields carry the same dimension, $\delta$. In a direct generalization, which does not involve any new techniques, one can obtain the result for two different $\varphi$'s, say $\varphi (t_2) \ \text{and} \ \tilde{\varphi} (t{_3})$, which carry different dimensions, $\delta \ \text{and} \ \tilde{\delta}$. The result for the 4-point function is
\begin{gather}
\widetilde{G}_4 \, (t_1, t_2, t_3, t_4) \equiv \brakets{t_1}{\varphi (t_2)\, \tilde{\varphi} (t_3)}{t_4} = \nonumber\\[1ex]
 \brakets{n=0}{\varphi (0)}{n=0} \,  \brakets{n=0}{\tilde{\varphi} (0)}{n=0} \frac{\Gamma (2 r_0)}{2^{\delta +\tilde{\delta} + 2 \, r_0}}\ \times \nonumber\\[1ex]
\frac{1}{(t_{13})^{\delta-r_0}\, 
(t_{24})^{\tilde{\delta}-r_0}\, 
(t_{12})^{\tilde{\delta}+r_0}\, 
(t_{34})^{\delta+r_0}\,
(t_{14})^{2r_0 - \delta - \tilde{\delta}}   }  \ \times \nonumber\\[1ex]
x^{r_0} \ _2 F_1 \ (\delta, \tilde{\delta}; 2 r_{0}; x ) .
\label{rjAdS2-14}
\end{gather} 

The conformal block $\widetilde{\mathit{B}}$ with the eignenvalue $\Delta (\Delta-1)$ remains as in \eqref{rjAdS2-12}. It matches \eqref{rjAdS2-14} when $\Delta = \Delta_1 = \Delta_4 = r_0, \Delta_2 = \delta, \Delta_3 = \tilde{\delta}$. When $\delta = \tilde{\delta}$, $\widetilde{G}\ \text{and}\  \widetilde{B}$ reduce to \eqref{rjAdS2-2} and \eqref{rjAdS2-13}.

\section{Various Observations on the Formalism}\label{rjSec3}
The construction of the states $\ket{t}$ in \eqref{rjAdS7}, \eqref{rjAdS8} has found response in the literature \cite{Nakayama:2011qh}. Therefore, we elaborate some of their further properties, which follow from \eqref{rjAdS2} and \eqref{rjAdS10}.

\subsection{Energy Eigenstates}
Since the action of $H$ on $\ket{t}$ is known form \eqref{rjAdS10-a}, it is readily seen that \cite{rjbibitem7}
\begin{equation}
\ket{E} = 2^{r_0}\, \frac{E^{1/2}}{(a E)^{r_0}}\ \int\limits^{\infty}_{-\infty}\, \frac{d t}{2\pi}\ e^{-i E t}\, \ket{t}
\label{rjAdS3-1}
\end{equation}
is an orthonormal energy eigenstate. The prefactor ensures normalization.
\begin{equation}
\braket{E}{E^\prime} = \delta (E - E^\prime)
\label{rjAdS3-2}
\end{equation}
The $SO (2,1)$ generators act as
\begin{subequations}\label{rjAdS3-3}
\begin{eqnarray}
H \ket{E} &=& E \ket{E},\label{rjAdS3-3a}\\[1ex]
D \ket{E} &=& i\, \left(E \frac{d}{d E} + \frac{1}{2}\right)\ \ket{E} , \label{rjAdS3-3b} \\[1ex]
K \ket{E} &=& \left(-E  \frac{d^2}{d E^2} -  \frac{d}{d E} + \frac{(r_0 -1/2)^2}{E}\right)\ \ket{E} . \label{rjAdS3-3c}
\end{eqnarray}
\end{subequations}
The $\ket{E}$ states allow establishing further properties of the $\ket{t}$ states, whose overlap with $\ket{E}$ is determined from \eqref{rjAdS13} and \eqref{rjAdS3-1}.
\begin{equation}
\braket{t}{E} = 2^{-r_0}\, \frac{(a E)^{r_0}}{E^{1/2}}\ e^{-i E t} 
\label{rjAdS3-4}
\end{equation}

\subsection{(In)-Completeness of the $\mathbf{\ket{t}}$ States}\label{rjSec3B}
Combining \eqref{rjAdS3-1} with \eqref{rjAdS3-4} gives
\begin{subequations}\label{rjAdS3-5}
\begin{equation}
\ket{E} = 2^{2 r_0}\, \frac{E}{(a E)^{2 r_0}}\ \int\limits^{\infty}_{-\infty}\, \frac{d t}{2\pi} \ \ket{t} \braket{t}{E} ,
\label{rjAdS3-5a}
\end{equation}
or
\begin{equation}
2^{-2 r_0}\ \frac{(a H)^{2 r_0}}{H}\ \ket{E} = \int\limits^{\infty}_{-\infty}\, \frac{d t}{2\pi} \ \ket{t} \braket{t}{E} .
\label{AdS3-5b}
\end{equation}
\end{subequations}
Since the energy eigenstates are complete, we arrive at an (in-)complete relation for the $\ket{t}$ states.
\begin{equation}
\frac{1}{H}\ \left(\frac{a H}{2}\right)^{2 r_0} =  \int\limits^{\infty}_{-\infty} \frac{d t}{2\pi} \, \ket{t} \bra{t}
\label{rjAdS3-6}
\end{equation}

\subsection{State-Operator Correspondence}\label{rjSec3C}
In the Letter \cite{Chamon:2011xk} it is shown that
\begin{equation}
\ket{\psi} \equiv e^{- H a}\,  \ket{t= 0}
\label{rjAdS3-7}
\end{equation}
satisfies $R \ket{\psi} = r_0 \ket{\psi}$;  hence $\ket{\psi}$ is proportional to  $\ket{n=0}$. Naming the proportionality constant $\mathcal{N}$, we have
\begin{subequations}\label{rjAdS3-8}
\begin{eqnarray}
&&\ket{\psi} = \mathcal{N}\, \ket{n=0} ,\nonumber\\
|\mathcal{N}|^2 &=& \braket{\psi}{\psi} = \brakets{t=0}{e^{-2 H a}}{t = 0} ,\nonumber\\[1ex]
&=& \int^{\infty}_{0} \, d E \, e^{-2 E a} \, |\braket{t=0}{E }|^2 .
\label{rjAdS3-8a}
\end{eqnarray}
The matrix element (with $a$ restored) is given by \eqref{rjAdS3-4}. Therefore
\begin{equation}
|\mathcal{N}|^2 =  \int^{\infty}_{0} \, d\, E \, e^{-2 E a}\, \frac{1}{E}\ \left(\frac{a E}{2}\right)^{2 r_0} = \frac{\Gamma (2 r_0)}{4^{2 r_0}} .
\label{rjAdS3-8b}
\end{equation}
\end{subequations}
Then \eqref{rjAdS3-7} and \eqref{rjAdS3-8} show that
\begin{subequations}\label{rjAdS3-9}
\begin{eqnarray}
e^{-H a}\, \ket{t=0} &=& \frac{1}{2^{2 r_0}}\ \Gamma^{1/2}\, (2 r_0) \, \ket{n=0} , \nonumber\\[1ex]
\ket{t=0} &=& \frac{1}{2^{2 r_0}}\ \Gamma^{1/2}\, (2 r_0) \, e^{H a} \ket{n=0} .
\label{rjAdS3-9a}
\end{eqnarray}
Since $H$ generates $t$-evolution, a further consequence is \cite{rjbibitem8}
\begin{equation}
\ket{t} = e^{i H t}\, \ket{t=0} = \frac{\Gamma^{1/2}\, (2 r_0)}{2^{2 r_0}}\ e^{(a + i t)\, H}\, \ket{n=0} .
\label{rjAdS3-9b}
\end{equation}
\end{subequations}
This is an interesting alternative to (\ref{rjAdS7}), \eqref{rjAdS8}.

\subsection{Alternative Evolution}\label{rjSec3D}
In our treatment evolution takes place  in $t$ time and is generated by $H$. This is seen in \eqref{rjAdS10-a} and \eqref{rjAdS11-a}, where the action of $H$ is time derivation, {\it i.e.} infinitesimal time translation.

However, our formalism is based on $R$, rather than $H$. Thus recasting evolution so that it is generated by $R$ becomes an interesting alternative. This is accomplished by redefining time $t$.

Observe from \eqref{rjAdS10} that
\begin{equation}
R \ket{t} = \frac{1}{2}\ \left(a H +\frac{K}{a}\right)\ \ket{t} = -i \left(\frac{1}{2} \ {[a +t^2/a]}\ \frac{d}{d t} + \frac{r_0 t}{a}\right)\ \ket{t} .
\label{rjAdS3-10}
\end{equation}
Upon defining a new ``time" $\tau$,
\begin{equation}
t = a \tan\ \tau/2
\label{rjAdS3-11}
\end{equation}
[compare \eqref{rjAdS7}] the expression in the last parenthesis of \eqref{rjAdS3-10} may be rewritten as 
\[
(\cos \tau/2)^{2 r_0}\ \frac{d}{d \tau}\ \bigg((\cos \tau/2)^{-2 r_0}\, \ket{t = a \tan \tau/2}\bigg) .
\]
Hence if we define new ``time"states $\ket{\tau}$
\begin{equation}
\ket{\tau} = (\cos \tau/2)^{-2 r_0} \ket{t = a \tan \tau/2},
\label{rjAdS3-12}
\end{equation}
it follows that $R$ translates $\tau$ infinitesimally.
\begin{equation}
R \ket{\tau} = - i\, \frac{d}{d \tau}\ \ket{\tau}
\label{rjAdS3-13}
\end{equation}
Explicitly the state $\ket{\tau}$ is given by
\begin{subequations}\label{rjAdS3-14}
\begin{eqnarray}
&& \ket{\tau} = \widetilde{N} (\tau) \exp - (e^{i \tau}\, L_+) \ket{n=0}, \label{rjAdS3-14a} \\
\widetilde{N} (\tau) &=& (\cos \tau/2)^{-2 r_0}\, N\, (t = a \tan \tau/2) , \nonumber \\
&=& \left[\Gamma (2 r_0)\right]^{1/2}\ e^{\, i r_0 \tau}. \label{rjAdS3-14b}
\end{eqnarray}
\end{subequations}
The spectrum of $H$ is continuous and the conjugate time variable is unrestricted. On the other hand, the spectrum of $R$ is discrete, equally spaced, and the conjugate $\tau$ variable is periodic.

In terms of the new variable, the 2-point function becomes \cite{rjbibitem8}
\begin{eqnarray}
G_2 \, (\tau^\prime, \tau) = \frac{\Gamma (2 r_0)}{\left[ 2 i \{\sin \left[\frac{\tau - \tau^\prime}{2}\right]
	\}               
 \right]^{2r_0}}\ .
 \label{rjAdS3-15}
\end{eqnarray}

One may also consider evolution generated by $\displaystyle{\frac{1}{2} \left(a H - \frac{K}{a}\right)}$. This development begins when the new time $\tau$ is defined as $t= a \tanh {\tau/2}$, which leads to similar replacement in \eqref{rjAdS3-11} -- \eqref{rjAdS3-15} of trigonometric functions by hyperbolic ones.

\section*{Conclusion}
We have studied the 4-point function and its conformal block for CFT$_1$ --- conformal quantum mechanics. We used operators that are not primary $[\mathcal{O} (t)]$ and states that are not invariant  $[R$-vacuum $\ket{n=0}]$.  Nevertheless results obey the conformal constraints. 

For the 2- and 3- point functions an AdS$_2$ bulk dual can be identified. \cite{Chamon:2011xk} We have not accomplished that for the 4-point function. But the simplicity of the block structure --- just one block is needed to reproduce the 4-point function --- gives the hope that a dual model in the AdS$_2$ bulk can be found. It is interesting to observe that the AdS$_2$ bulk propagator is given by a hypergeometric function, just as $G_4$ and its conformal block

We acknowledge  conversations with S. Behbahani , C. Chamon, D. Harlow and L. Santos. This research is supported by DOE grants DE-FG02-05ER41360 (RJ) and DE-FG02- 91ER40676 (SYP).


\begin{thebibliography}{99}
\bibitem{rjbibitem3a}
A careful evaluation shows that the singularity in $G_2(t_1, t_2) \ \text{at}\ t_1 =t_2$ is regulated as $t_1 - t_2 \to t_1-t_2 - i\, \varepsilon$.
\end{thebibliography}
\end{document}